\documentclass[aps, prb, showpacs, 10pt, floatfix, groupedaddress]{revtex4-1}

\usepackage{amssymb,amsmath,xcolor,graphicx,xspace,colortbl,rotating} 
\usepackage{amssymb}  
\usepackage{amssymb,amsmath,xcolor,graphicx,xspace,colortbl,rotating}  
\usepackage{bm}  
\usepackage{color}  
\usepackage{graphicx}  
\usepackage{times}  
\graphicspath{{Symboli2Hofstadter_graphics/}{Symboli2Hofstadter_tcache/}{Symboli2Hofstadter_gcache/}}
\DeclareGraphicsExtensions{.pdf,.eps,.ps,.png,.jpg,.jpeg}
\graphicspath{{Symboli2Hofstadter_graphics/}{Symboli2Hofstadter_tcache/}{Symboli2Hofstadter_gcache/}}
\DeclareGraphicsExtensions{.pdf,.eps,.ps,.png,.jpg,.jpeg}
\begin{document}
\title{A higher-dimensional quasicrystalline approach to the Hofstadter and Fibonacci butterflies topological phase diagram and band conductance:  symbolic sequences, Sturmian coding and self-similar rules at all magnetic fluxes}
\author{Gerardo Naumis}
\affiliation{Departamento de Sistemas Complejos, Instituto de F\'{i}sica, Universidad Nacional Aut{\'o}noma de M{\'e}xico (UNAM), Apartado
Postal 20-364, 01000 M{\'e}xico, Distrito Federal, M{\'e}xico}
\begin{abstract}The topological properties of the quantum Hall effect in a crystalline lattice, described by Chern numbers of the Hofstadter butterfly quantum phase diagram, are 
deduced by using a geometrical method to generate the structure of quasicrystals: the cut and projection method. Based on this, we provide a geometric unified approach to the Hofstadter topological phase diagram at all fluxes. Then we show that for any flux, the bands conductance follow a two letter symbolic sequence . As a result, bands conductance  at different fluxes obey inflation/deflation rules as the ones observed to build quasicrystals. 
The  bands conductance symbolic sequences are given by the Sturmian coding of the flux and can be found by considering a circle map, a billiard or trajectories on a torus. Simple and fast techniques are thus provided to obtain Chern numbers at any magnetic flux.  This approach rationalize the previously observed topological equivalences between the Fibonacci and Harper potentials (also known as the almost Mathieu operator problem) or with other trigonometric potential, as well as the relationship with Farey sequences and trees.
\end{abstract}
\maketitle

\section{Introduction}
Historically, the quantum Hall effect (QHE) was the first discovered manifestation of a topological phase \cite{TKKN}. Before, the spectrum as a function of the magnetic flux  was firstly found by D. Hofstadter\cite{Hof}.
As seen in Fig. \ref{Fig:Butterfly}, this spectrum is a beautiful fractal which was so-called the Hofstadter butterfly. It has been measured using different kind of effective systems, but only recently it has been possible to  measure it in atomic systems \cite{Dean2013}. In this moment, there is a huge interest in this problem, as there is a connection yet no understood between the solutions of the QHE and the superconductance for Moir\'e patterns at magic angles made from graphene over graphene \cite{Tarnopolsky2019}.    

Also, the interest on the Hofstadter
butterfly has been growing in the context of topological insulators and two-dimensional materials \cite{TI,KM2,Taboada,Naumisreview,Karnaukhov,Rami}. These insulators are exotic states of matter which are insulators in the bulk but conduct
along the edges\cite{TKKN,Fradkin,TI}. They are characterized by topologically protected
gapless boundary modes, known as \textit{edge-Chern} modes. These modes manifest the nontrivial band structure topology of the bulk\cite{TKKN}
and their number equals the topological integer known as the Chern number
($\sigma_r$). These Chern numbers are the quanta of the Hall conductance for a system under a constant magnetic field \cite{TKKN}.
Each Chern quantum number is thus associated to a gap $r$. The Chern number for each gap is obtained by solving a Diophantine equation \cite{Fradkin}. 
Recently, there have been many works to find the Hofstadter butterfly phases for other systems \cite{Square,ThreeBand}, usually related with graphene \cite{Igor1,Karnaukhov}.

There is a vast amount of literature dedicated to the subject (the original paper by D. Hofstadter has more than
$4,000$ citations),  most of it usually looks for scaling properties for a given flux or by looking at replicas of the Landau states and not for the whole fractal, although now there is a growing interest in the global fractality and its relationship with other fractals \cite{Indubook,InduEPJ}. 
Moreover, the Hofstadter butterfly, obtained form the Harper equation \cite{Harper,Ostlund,Dana}  (which is also known as the almost Mathieu operator problem in mathemathics), is considered as one of the first examples of a quasiperiodic Hamiltonian, yet many works treat the QHE with different methodolgies than the ones used to describe quasicrystals \cite{QC,Steurer}. In a previous paper, we showed that the Harper potential and the Fibonacci chain were just examples of different kinds of trigonometric potentials  \cite{NaumisPHYSB} . Then one
can follow the transformation of the Harper model into the Fibonacci one just by adding harmonics to the potential, leading to a "Fibonacci butterfly" made from a square well potential \cite{NaumisPHYSB}.

Later on, our work was extended by  Kraus and  Zilberberg to show that in fact, the Harper model and the Fibonacci chain are within the same topological class \cite{HarperFib}.
An explicit experimental demonstration of transport, mediated by the edge mode was shown by an experiment by pumping light across the QC \cite{QCti}. Key to the topological characterization of Quasicrystals is the translational invariance that shifts the origin of a quasiperiodic system \cite{QCti,InduPRL}. Also, a novel manifestation of the topology that is
unique to QCs has been found, since \textit{band edge modes} encode topological invariants in their spatial profiles \cite{InduPRL}.

\begin{figure*} \label{Fig:Butterfly}
\includegraphics [width =0.8\linewidth,height=0.8 \linewidth ]{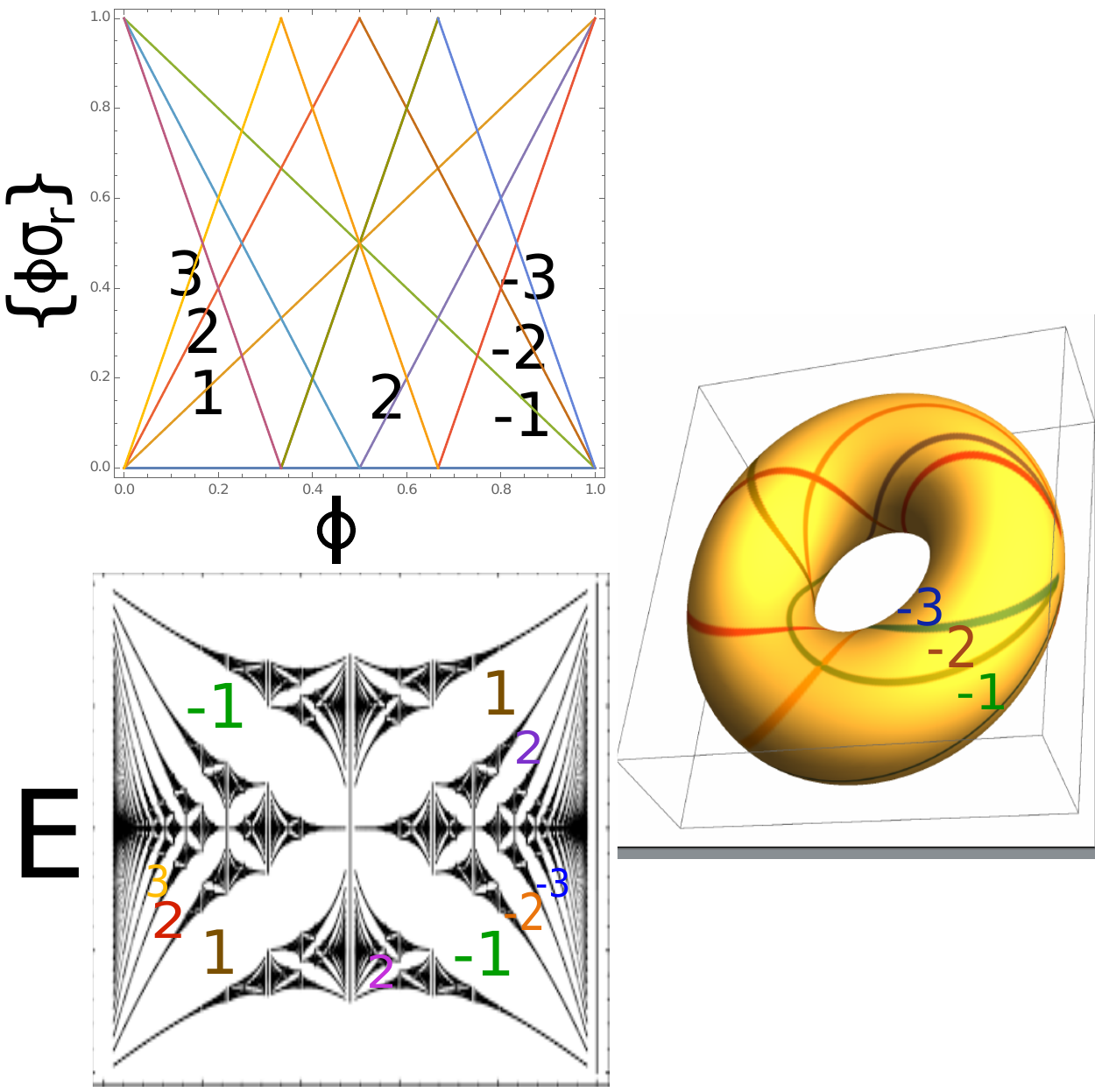}
\leavevmode 
\caption{On top left, topological map of the Hofstadter butterfly for the first Chern numbers, i.e., the filling fractions $r/q=\{\phi \sigma_r\}$ for  $\sigma_{r}=\pm 1,\pm 2, \pm 3$ as a function of $\phi$. Bottom left, the map is compared with the Hofstadter butterfly where each gap has a Chern number associated with its conductance. Each gap associates with a line in the map. To the right, the topological map of the Hofstadter butterfly on a torus. This map is obtained by defining two angles $\Phi=2\pi \phi$ and  $\theta=2 \pi \{\sigma_r \phi\}$. Here $\Phi$  is the azimuth angle, also known as the "toroidal" direction, and $\theta$ is the "poloidal" angle. For each Chern number $\sigma_r$, a trajectory on the torus is obtained, represented here with a different color and with its corresponding label. Note how trajectories crossings produce topological sequences nearby Van Hove singularities\cite{Naumis}.}
\end{figure*}

This article continues the search for a common lenguage to encode topological properties and quasicrystals. First we formalize the relationship the topological properties of the Hofstadter and Fibonacci butterflies using a classic cut and projection quasicrystallographic description \cite{Dana,Levine,Naumis2005}.  The second propose is to show how this approach allows to 
relate symbolic sequences to band conductances and then explain the relationship between electron diffraction and the topological phases.

The layout of this work is the following. In section \ref{High} we revise a method developed previously by the author to find the Chern numbers.
In section \ref{Cut}  the method is written in terms  of the cut and projection method, while section \ref{Conductances}  is devoted to find the conductance as symbolic sequences. As shown in section \ref{Recursion}, this allows to obtain simple methods to find Chern numbers. 
In section \ref{diffraction} we relate  the cut and projection method with the Harper potential  properties, and finally, the conclusions are given.

\section{ Topological Phase diagram: Higher-Dimensional approach}\label{High}

In this section, we will consider some general properties of the Hoftstadter butterfly topological map. As shown by the author before, this topological phase diagram can be made by using a higher dimensional approach \cite{Naumis}. Here we outline the main results to introduce the connection with the cut and projection method.  First we observe that the Hoftstadter spectrum (see Fig. \label{Fig:Butterfly}) is produced from the Harper equation \cite{Ostlund},
\begin{equation}\label{EqHarper}
 \psi_{m+1}^{r}+\psi_{m-1}^{r}+V(m)\psi_{m}^{r}=E_{r} \psi_{m}^{r}
\end{equation}
where $\psi_{m}^{r}$ are the electron wavefunctions at site $m$ for the band $r$ with energies $E_r$. The Harper potential is\cite{Ostlund},
\begin{equation}
V(m)=2\lambda \cos (2\pi m \phi +2\pi \nu_y)
\end{equation}
and $0 \leq \nu_{y} \leq 1/2$ for $p$ odd and $0 \leq \nu_{y} \leq q/2p$ for $p$ even\cite{TI}.

\begin{figure*}
	\includegraphics [width =0.4\linewidth ,height=0.4\linewidth]{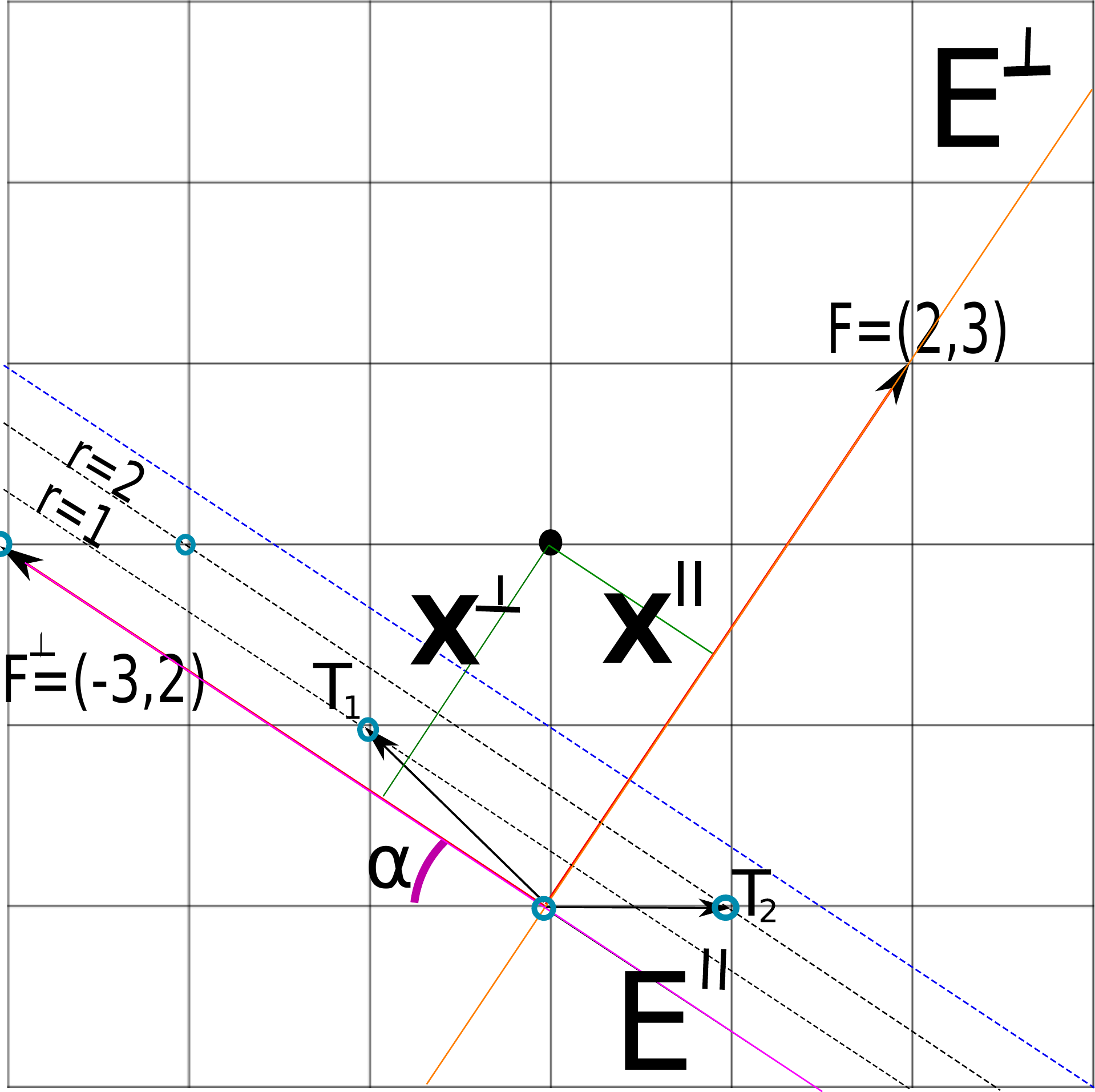}
	\leavevmode  \caption {Cut and projection method applied to find the solutions of the Diophantine equation given by Eq. (\ref{Diophantine}). Here, the flux vector is chosen to give $\phi=\tan \alpha =2/3$.The set of parallel lines gives possible solutions for $\boldsymbol{F}(\phi)\cdot \boldsymbol{T}_r=r$ for each gap $r$. For example, $\bf{T}_1$ and $\bf{T}_2$ are solutions for $r=1$ and $r=2$, others are indicated by open circles. Notice the periodicity of the solutions. In general, the flux vector $F$ defines a parallel subspace $E^{\vert \vert}$, while $E^{\perp}$ is a perpendicular subspace defined by $F^{\perp}$. Any point $\boldsymbol{X}$, indicated by a black dot, can be decomposed as $\boldsymbol{X}=\boldsymbol{X}^{\vert \vert}+\boldsymbol{X}^{\perp}$. Valid solutions require $\boldsymbol{X}$ to have integer coordinates and $|\boldsymbol{X}^{\perp}|<1$. Physically, $\boldsymbol{X}^{\perp}$ is the band filling ratio $r/q$.}
	\label{Fig:Cut}
\end{figure*}

The energy $E$ as a function of the flux $\phi $ produces the Hofstadter butterfly shown in Fig. \ref{Fig:Butterfly}. For a  given flux $\phi  =p/q$, a Chern number $\sigma _{r\;\text{}\;}$ is associated with the gap $r$, counted from the bottom to the top of the spectrum. The Chern number gives the conductance of such gap \cite{TKKN}. The gap and its corresponding Chern number is obtained by solving the following Diophantine equation \cite{TKKN,Fradkin,HK}, 

\begin{equation}\label{Diophantine}
r =p \sigma _{r} +q \tau _{r}
\end{equation}
where $\tau _{r}$ is an integer.  To solve this equation we can go to a higher dimension as follows\cite{Naumis}. As seen in Fig. \ref{Fig:Cut}, define a flux vector , 
\begin{equation}
\boldsymbol{F}(\phi)=(p,q)
\end{equation}
and a topology vector, 
\begin{equation}
\boldsymbol{T}_r=(\sigma_r,\tau_r)
\end{equation}

In this lenguage, the Diophantine is written as, 

\begin{equation}
r=\boldsymbol{F}(\phi) \cdot \boldsymbol{T}_r
\label{diophantine}
\end{equation}

Thus the gap index $r$ is the projection of the topology vector onto the flux vector. This is no other than the distance between the point $(\sigma _{r} ,\tau _{r})$ and a line perpendicular to $\boldsymbol{F}(\phi)$.

This suggest a method to solve the Diophantine equation. First we take a 2D space, in which any point is denoted as $\boldsymbol{X}$. As seen in Fig. \ref{Fig:Cut}, we consider a vectorial subspace of lower dimensionality $E^{\vert \vert }$, in this case a line perpendicular to $\mathbf{F}(\phi)$. Points in this line have the form $\boldsymbol{X} \cdot \boldsymbol{F}(\phi)=0$.

A perpendicular subspace $E^{ \perp }$ is now defined, as indicated in Fig. \ref{Fig:Cut}. All possible solutions to the Diophantine are contained in the family of parallel lines  $\boldsymbol{X} \cdot \boldsymbol{F}(\phi)=r$. This is equivalent to find all integer coordinates that are within the parallel lines  $\boldsymbol{X} \cdot \boldsymbol{F}(\phi)=0$ and $\boldsymbol{X} \cdot \boldsymbol{F}(\phi)=q$. We will call this region as the "band".

To find the solution of the Diophanitine equation we must proceed as in the cut and projection
method. The steps are the following, 

1) Consider a square lattice in 2D, such that $\boldsymbol{X} =(n_1 ,n_2)$ with $n_1,n_2$ any integer. 

2) Choose points such that $\boldsymbol{X}  \cdot \boldsymbol{F}(\phi) \leq q$. 

3) Then identify $n_1 =\sigma _{r}$ and $n_2 =\tau _{r}$. It is easy to show that integer coordinates points  $\boldsymbol{X}$  within the band satisfy \cite{Naumis},

\begin{equation}
\tau _{r} = -\lfloor \phi  \sigma _{r}\rfloor 
\end{equation}

where $\lfloor z\rfloor $ denotes the floor function of $z$. The floor function allows to select points $\boldsymbol{X}$ that fall inside the band. Thus gaps are labeled by the coordinates of a two dimensional lattice,
\begin{equation}\label{EqPhasePoint}
(\sigma _{r} ,\tau _{r}) =(\sigma _{r} , -\lfloor \phi  \sigma _{r}\rfloor )
\end{equation}

By using that any number $z$ can be written as $z =\lfloor z\rfloor  +\{z\}$, where $\{z\}$ denotes the fractional part of $z$ (observe that a negative number -$x$, we have $\{ -x\} =1 -\{x\}$), we can express $\tau _{r}$ as, 

\begin{equation}\label{map}
r =q \{\phi  \sigma _{r}\} 
\end{equation}

Eq. (\ref{map}) can be inverted using the same methodology giving the Chern numbers as a function of the gap index, 

\begin{equation}\sigma _{r} = \left(\frac{q}{2} -q \{\phi  r +\frac{1}{2}\}\right)\zeta \label{invmap}
\end{equation}
where $\zeta=(-1)^{q-p}$ determines the correct sign in order to have a positive band index $r$. We will refer to these two previous equations as the hull functions. Several properties are deduced from Eq. (\ref{map}) and Eq. (\ref{invmap}). For rational $\phi$,

\begin{enumerate}
\item The solutions are periodic up to a vector $( -q ,p)$ , i.e., Cherns numbers have a period $q$ while $\tau_{r}$ has period $p$. 

\item The solutions for conductance correspond to Cherns between $ -q/2$ and $q/2$. This defines a ``first Brillouin zone'' for Cherns. 

\item The solution $\boldsymbol{T}_1$ for $r=1$ always exists, since a Diophantine equation of the form $a x +b y =1$ always has solution if $p$ and $q$ are relative primes. 

\item Then all solutions are obtained from the $r =1$ solution. To show this, consider the solution for $r =1$. It satisfies,
\begin{equation}
 \boldsymbol{F}(\phi) \cdot \boldsymbol{T}_1 =1
\end{equation}
multiplying this equation by $r$, it will satisfy the Diophanite equation Eq. (\ref{diophantine}). Then,
\begin{equation}
 \boldsymbol{T}_r  =r\boldsymbol{T}_1
\end{equation}

\item Combining the previous properties, the solutions are given by,
\begin{equation}
\bm{T}_r=r\bm{T}_1+s\bm{F}^\perp
\end{equation}
where $s$ is chosen to have Cherns between $ -q/2$ and $q/2$. This is equivalent to take solutions modulus $q$ in $\sigma _{r}$ and modulus $p$ in $\tau _{r}$.

\end{enumerate}

If we think Eq. (\ref{map}) as a function of $\phi$ for each integer $\sigma_r$, we obtain the Claro-Wannier map \cite{CW} 
  seen in Fig. \ref{Fig:Butterfly} a), which can be compared with the original butterfly  \ref{Fig:Butterfly} b). Each line corresponds to a gap and the slope of the line gives the Chern number.  Fig. \ref{Fig:Butterfly} b) the Chern labeling on the butterfly \cite{Naumis}.
The sawtooth function $\{\sigma_r \phi \}$ has period $1/\sigma_r$, and thus can be used to warp a torus for each Chern number $\sigma_r$. So consider the map $\Phi=2\pi \phi$ and 
$\theta=2 \pi {\sigma_r \phi}$ as a parametrization of the torus, in which $\Phi$  is the azimuth angle, known as the "toroidal" direction, and $\theta$ is the "poloidal" angle. In Fig. \ref{Fig:Butterfly} c)  we present the trajectories on the tours for first Chern numbers. Notice how Fig. \ref{Fig:Butterfly} c) is obtained by projecting the Claro-Wannier diagram of Fig. \ref{Fig:Butterfly}  onto a torus. It is interesting to observe that trajectories crossings corresponding to Van Hove singularities existing at all band centers due to saddle points of the energy dispersion \cite{Naumis}. 

\section{Cut and projection: structure of quasicrystals and topological phases}\label{Cut} 

The method exposed in the previous section turns out to be the same as one of the used to generate the structure of quasicrystals: the cut and projection method \cite{Steurer}.   
For further reference, let us now revist this method. To build the structure of a quasicristal,  
consider points $\boldsymbol{X}$ in  a $D$ dimensional space periodic lattice,
\begin{equation}
 \boldsymbol{X}=\sum _{j =1}^{D}n_{j}\boldsymbol{\hat{e}}_{j}
\end{equation}
where $\boldsymbol{\hat{e}}_{j}$ are the lattice vectors of a hypercubic ($D>3$), cubic ($D=3$) or square lattice ($D=2$).
These lattice points are projected onto a subspace $E^{\vert \vert }$ using a projection operator $\hat{\Pi} (\boldsymbol{X})$. This projection will be called $\boldsymbol{X}^{\vert \vert }$. A perpendicular subspace  $E^{\perp}$ to $E^{\vert \vert }$ is now defined. Any point $\boldsymbol{X}$ is decomposed as $\boldsymbol{X}=\boldsymbol{X}^{\vert \vert }+\boldsymbol{X}^{\perp}$, where $\boldsymbol{X}^{\perp}$ is the projection onto $E^{\perp}$.
Not all points $\boldsymbol{X}$ are selected to build the quasicrystal. Instead, points $\boldsymbol{X}$ are selected by using 
a band  function  $W(\boldsymbol{X}^{\perp})$ such that an acceptance width is given in the $E^{\vert \vert }$ space, 
resulting in,
\begin{equation}\label{Eq:QCinD}
\boldsymbol{R} =\hat{ \Pi} (\boldsymbol{X})W(\boldsymbol{X}^{\perp })=\boldsymbol{X}^{\vert \vert }W(\boldsymbol{X}^{\perp })
\end{equation}
where,
\begin{equation}\label{Eq:BandDef}
W(\boldsymbol{X}^{\perp }) =\left \{\begin{array}{c}1\text{ if
}   |\boldsymbol{X}^{\perp }|<1\\
0\text{  if
} |\boldsymbol{X}^{\perp }| \geq 1 \end{array}\right .
\end{equation}
Since the points $\boldsymbol{X}$ form a  lattice in D dimensions, using the linearity of the operator, 
it easy to prove that points in the quasicrystal are given by,
\begin{equation}\boldsymbol{R }=\left(\sum _{j =1}^{D}n_{j}\boldsymbol{q}_{j} \right) W(n_{1} ,n_{2} , . . . ,n_{D})
\end{equation}
where $n_{j}$ are integers and $\boldsymbol{q}_{j}$ is the projection of the higher-dimensionality base into $E^{\vert \vert }$, i.e., $\boldsymbol{q}_{j} =\hat{\Pi} (\boldsymbol{\hat{e}}_{j})$.

Let us now use this method to build one dimensional quasicrystals and rational approximants. As explained in Fig. \ref{Fig:Cut}, we first consider a  square-lattice. The subspace $E^{\vert \vert }$ is now a line inclined with angle $-\alpha$, while $E^{\perp }$ is a line perpendicular to it.

The points $\boldsymbol{X}$ in 2D with integer coordinates have the form $\boldsymbol{X} \equiv \boldsymbol{X}_{n_1,n_2}=(n_1,n_2)$. The
projection in $E^{\vert \vert }$ is given by,
\begin{equation}
  \boldsymbol{X}^{\vert \vert }\equiv \boldsymbol{X}_{n_1,n_2}^{\vert \vert }=n_1 q_1-n_2 q_2
\end{equation}
where,
\begin{equation}
q_{1}=\cos \alpha= \frac{q}{\sqrt{p^{2}+q^{2}}}; \
q_{2}= \sin \alpha=\frac{p}{\sqrt{p^{2}+q^{2}}}  
\end{equation}
and the perpendicular, 
\begin{equation}
\boldsymbol{X}^{\perp}\equiv \boldsymbol{X}_{n_1,n_2}^{\perp}=n_1 q_2+n_2 q_1
\end{equation}
 From this, the band condition (\ref{Eq:BandDef}) results here in a relationship between $n_1$ and $n_2$, to give $W(\boldsymbol{X}^{\perp})=\delta_{n_1,-\lfloor  n_{1} \tan \alpha \rfloor}$, with $\delta_{ij}$ is the Kronecker delta of $i$ and $j$. Finally, using the projection of the basis vectors $\boldsymbol{\hat{e}}_{1}=(1,0)$ and $\boldsymbol{\hat{e}}_{2}=(0,1)$ into the line $E^{\vert \vert }$, we obtain the positions along the sequence,
\begin{equation}
 R_{n_1} \equiv \boldsymbol{X}_{n_1,n_2}^{\vert \vert }W(\boldsymbol{X}_{n_1,n_2}^{\perp})=n_1 q_{1}+\lfloor n_1 \tan \alpha \rfloor q_{2}
\end{equation}

For irrational $\tan \alpha$, the sequence is quasiperiodic. The famous Fibonacci chain is obtained by using $\tan \alpha = \tau^{-1}$, where $\tau^{-1}=(\sqrt{5}-1)/2$ is the inverse golden mean. This method can be adapted to generate quasicrystals in 2D and 3D by using apropiate analytical expressions for the window function \cite{NaumisAragon1,NaumisAragon}. 

It is worthwhile mentioning that $R$ can be written as an average periodic chain, plus a flutuation part. Using the identity $x=\lfloor x \rfloor+\{x\}$
\begin{equation}
 R_{n_1}=n_1  <q>-\{ n_1 \tan \alpha  \}q_{2}
\end{equation}
where $<q>=q_{1}+\tan \alpha q_{2}$ is an average lattice parameter and the fractional part is the fluctuation part. The distances between consecutive points is given by,
\begin{equation}\label{EqDeltaR}
 |\delta R_{n_1}|=[\{ (n_1+1) \tan \alpha  \}-\{ n_1 \tan \alpha  \}]q_{2}
\end{equation}

Notice that other approximants or quasicrystals in the same local isomorphism class can be obtained by performing a translation of the width 
function along $E^{\perp }$.  These extra degrees of freedom are known as phasons, which are related with the extra phases that appear in the Fourier transform when
compared with a normal crystal. If the shift along $E^{\perp }$ is $\kappa$, then the sequence is transformed into,
\begin{equation}
 R_{n_1}=n_1 q_{1}+\lfloor n_1 \tan \alpha +\kappa \rfloor q_{2}
\end{equation}
or written as an average plus a fluctuation, 
\begin{equation}\label{positions}
  R_{n_1}=\kappa q_{2}+ n_1  <q>-\{ n_1 \tan \alpha +\kappa\}q_{2}
\end{equation} 
which shows that the effect is a shift of the origin. 

Now we can see how the topological phases of the Hofstadter butterfly are determined by the same method used to build quasicrystal. We set 
$\tan \alpha=\phi$  
and consider a higher-dimensional point $\boldsymbol{X}=\boldsymbol{X}_{\sigma_r,\tau_r}=(\sigma_r,\tau_r)$ representing a possible topological phase.  
The distance between this topological phase point and the line $E^{||}$ is given by,
\begin{equation}
|\boldsymbol{X}_{\sigma_r,\tau_r}^{\perp}|=\frac{\boldsymbol{X}_{\sigma_r,\tau_r}  \cdot \boldsymbol{F}(\phi)}{|\boldsymbol{F}(\phi)|}
\end{equation}
and by using Eq. (\ref{EqPhasePoint}), we obtain,
\begin{equation}\label{EqXperpFilling}
 |\boldsymbol{X}_{\sigma_r,\tau_r}^{\perp}|=\frac{r}{\sqrt{p^{2}+q^{2}}}=\frac{q}{\sqrt{p^{2}+q^{2}}}\{ \sigma_r \phi \}
\end{equation}
Thus  $|\boldsymbol{X}_{\sigma_r,\tau_r}^{\perp}|$ determines the filling fraction $r/q$.  From the previous equation, is clear a deep connection between the method to build quasicrystals  and topological phases.  We will explore such connections in the forthcoming sections,

\section{Band conductance as a symbolic sequence}\label{Conductances} 

 Let us first explain how the conductance is related with symbolic sequences akin to the structure of quasicrystals and its rational approximants. 
 In general, the contribution of a band $r$  to the conductance  is given by the difference between the Chern numbers associated with each band edge \cite{Fradkin},
 \begin{equation}\label{EqSigma}
  \sigma _{B} (r) =(\sigma _{r +1} -\sigma _{r})
 \end{equation}
where here the band and gap conductance is measured in units of $\frac{e}{h}$.
By using Eq. (\ref{invmap}) in the previous definition, we obtain that,
\begin{equation}\label{Eqbands}
\sigma _{B} (r) =q\left (\{\phi  r +\frac{1}{2}\} -\{\phi  (r +1) +\frac{1}{2}\}\right )
\end{equation} 
This is precisely the distance between consecutive points in a sequence obtained from the cut and projection methods as in Eq. (\ref{positions}), i.e., is the set of distances between
points in a rational approximant or in a quasicrystal.
To see this, observe that the function $\{x\}$ has the property  $\{a +b\} =\{a\} +\{b\}$ if  $\{a\} +\{b\} <1$ and  $\{a +b\} =1-(\{a\} +\{b\})$ if  $\{a\} +\{b\} >1$. Thus, it turns out 
that $\sigma _{B} (r)$ only takes two values, $ -p$ and $ -q$. We map these two values to  the letters $L$ and $S$. For $\zeta=1$,i.e., $q-p$ odd,
\begin{equation}-p \rightarrow  S
\end{equation}
\begin{equation}(q-p) \rightarrow L
\end{equation}
while  for $\zeta=-1$, i.e., $q-p$ even,
\begin{equation}
 p \rightarrow  S
\end{equation}
\begin{equation}
(p-q) \rightarrow L
\end{equation}

\begin{figure*}
	\includegraphics [width =0.6\linewidth ,height=0.6\linewidth ]{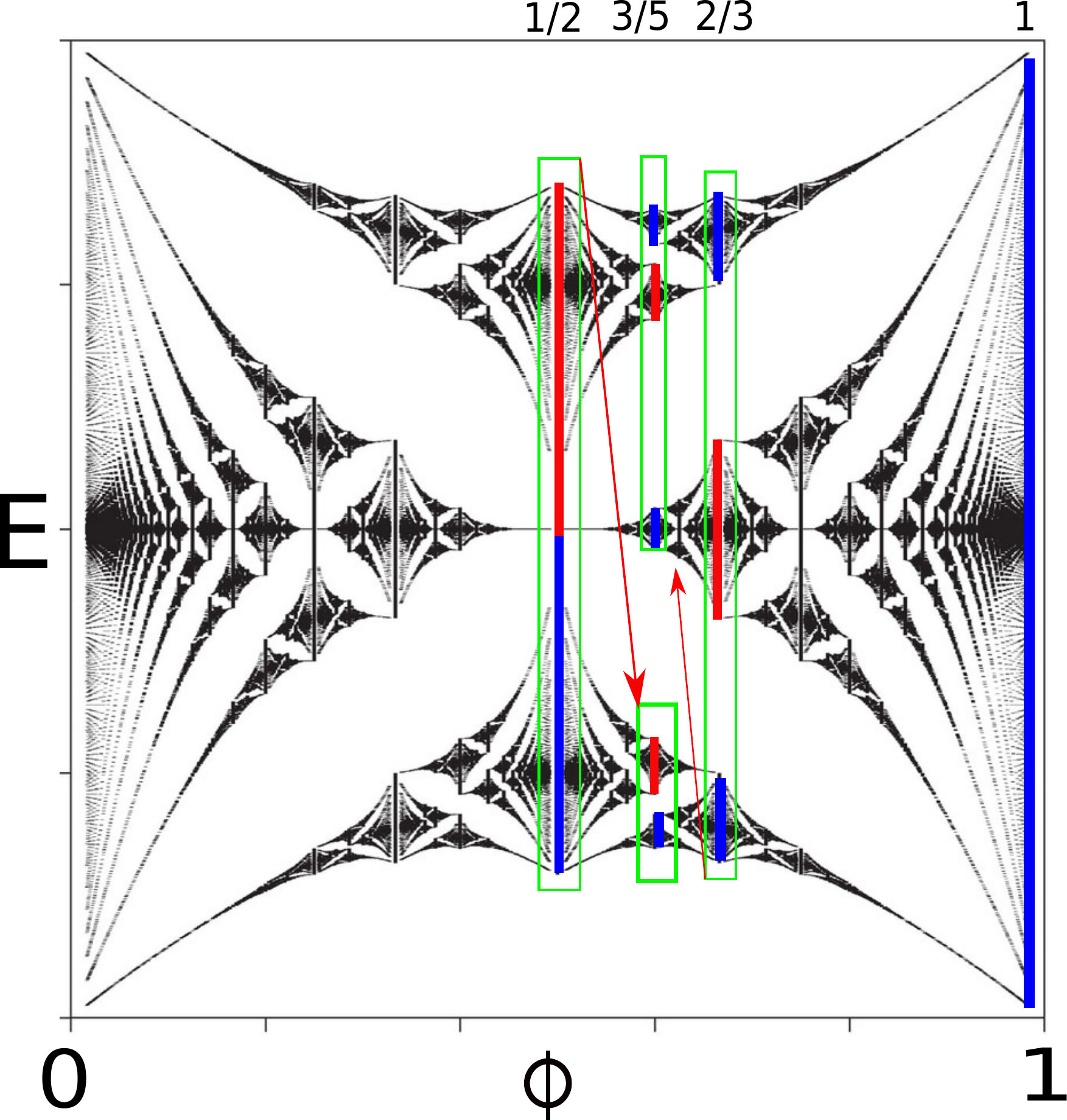}
	\leavevmode  \caption {Some symbolic sequences for bands conductance given in table 1, over imposed on the Hofstadter butterfly. Bands with index L are indicated in blue, while bands with S are in red. The sequence for $\phi=3/5$ is obtained by joining the sequences $\phi=1/2$ and $\phi=2/3$, as indicated at the figure top and by the green boxes.  Notice the inflation/deflation rules and the scaling of bands.} 
	\label{SymbolicButterfly}
\end{figure*}

 In Fig. \ref{SymbolicButterfly} we show some sequences on the original Hofstadter butterfly. For each rational $\phi $, the periodicity of the sequence is given by $q$. In fact, by comparing Eqns. (\ref{EqSigma}) and (\ref{EqDeltaR}) and setting $\kappa=1/2$,
we just proved that the band conductance is proportional to the fluctuation part of the sequence,
\begin{equation}
\sigma_{B}(r)= \delta R_{r+1}
\end{equation}

To further understand the previous results, let us denote the band conductance sequences for a given $\phi$ as $S_B(\phi)$.  In table 1 we show for several fluxes, the gap index $r$ and its associated Chern number $\sigma_r$, as well as the band conductances and the corresponding symbolic sequence. In these examples, each flux was chosen to  match the first rational approximant of the golden mean $(\sqrt{5}-1)/2$, given by the ratio of two successive Fibonacci numbers $\phi_j=F(j-1)/F(j)$. The $j$-esim Fibonacci number is given by $F(j)=F(j-2)+F(j-1)$, with $F(0)=1$ and $F(1)=2$.

\begin{table}[ht]
\begin{tabular}{ |p{1cm}||p{1cm}|p{1cm}|}
 \hline
 \multicolumn{3}{|c|}{$\phi=1$  \ ($\zeta=1$)} \\
 \hline
 $r$ & $0$ & $1$\\
 \hline
 $\sigma_r$ & $0$ & $1$\\
 \hline
 $\sigma_B(r)$ &$1$ & $-$ \\
 \hline
 $S(1)$& L  & -\\
 \hline
\end{tabular}

\begin{tabular}{ |p{1cm}||p{1cm}|p{1cm}|p{1cm}|p{1cm}|  }
 \hline
 \multicolumn{4}{|c|}{$\phi=1/2$  \ ($\zeta=-1$)} \\
 \hline
 $r$ & $0$ & $1$ & $2$ \\
 \hline
 $\sigma_r$ & $0$ & $-1$ &   $0$ \\
 \hline
 $\sigma_B(r)$ &$-1$ & $1$ & - \\
 \hline
 $S(1/2)$& L  & S &  -\\
 \hline
\end{tabular}

\begin{tabular}{ |p{1cm}||p{1cm}|p{1cm}|p{1cm}|p{1cm}|  }
 \hline
 \multicolumn{5}{|c|}{$\phi=2/3$ \ ($\zeta=-1$)} \\
 \hline
 $r$ & $0$ & $1$ & $2$ & $3$\\
 \hline
 $\sigma_r$ & $0$ & $-1$ &   $1$ & 0\\
 \hline
 $\sigma_B(r)$ &$-1$ & $2$ & $-1$ & -\\
 \hline
 $S(2/3)$& L  & S & L & -\\
 \hline
\end{tabular}

\begin{tabular}{ |p{1cm}||p{1cm}|p{1cm}|p{1cm}|p{1cm}|p{1cm}|p{1cm}| p{1cm}|}
 \hline
 \multicolumn{7}{|c|}{$\phi=3/5$  \ ($\zeta=1$)} \\
 \hline
 $r$ & $0$ & $1$ & $2$ & $3$ & $4$ & $5$ \\
 \hline
 $\sigma_r$ & $0$ & $2$ & $-1$ & $1$ & $-2$ & $0$\\
 \hline
 $\sigma_B(r)$ & $2$ & $-3$ & $2$ &  $-3$ & $2$ & $-$\\
 \hline
 $S(3/5)$& L  & S & L & S & L & $-$\\
 \hline
\end{tabular}

\begin{tabular}{ |p{1cm}||p{1cm}|p{1cm}|p{1cm}|p{1cm}|p{1cm}|p{1cm}|p{1cm}|p{1cm}|p{1cm}|}
 \hline
 \multicolumn{10}{|c|}{$\phi=5/8$ \ ($\zeta=-1$)} \\
 \hline
 $r$ & $0$ & $1$ & $2$ & $3$ & $4$ & $5$ & $6$ & $7$ & $8$ \\
 \hline
 $\sigma_r$ & $0$ & $-3$ & $2$ & $-1$ & $-4$ & $1$ & $-2$ & $3$ & $0$ \\
 \hline
 $\sigma_B(r)$ & $-3$ & $5$ & $-3$ &  $-3$ & $5$ & $-3$ & $-5$ & $-3$ & $-$\\
 \hline
 $S(5/8)$& L  & S & L & L & S & L & S & L & -\\
 \hline
\end{tabular}
\caption{Gap number $r$, the associated Chern number $\sigma_r$, the band conductance given by $\sigma_B(r)=\sigma_{r+1}-\sigma_r$ and the associated symbolic sequence $S(\phi)$ for fluxes chosen as the first golden mean approximants. Notice how a given symbolic sequence is given by joining the previous two sequences. Such construction
is seen in Figures \ref{SymbolicButterfly} and \ref{FibonacciButer}, where bands  conductances follow the same pattern. Also observe that sequences are similar to the usual Fibonacci ones up to a global phason due to the factor $1/2$ that appears in Eq. (\ref{EqSigma}). Here all sequences are the same as in Fibonacci except for $\phi=3/5$.}
\end{table}

As predicted by Eq. (\ref{Eqbands}), a symbolic sequence is obtained for the band conductances. Moreover, we observe that in fact, the sequences for different fluxes also follow a recursive relation similar to the used for Fibonacci chains, i.e., form Table 1 we see that,
\begin{equation}
 S(\phi_j)=S(\phi_j-2) \oplus S(\phi_j-1) 
\end{equation}
where the sign $\oplus$ means join two sequences. From example,  $S(5/8)=S(2/3) \oplus S(3/5)$. 
Although superficially this seems to be a Fibonacci sequence, in fact is very important to remark that the order of  joining chains $S(\phi_j-2) \oplus S(\phi_j-1)$ is reversed when compared to the usual Fibonacci chain in which $S(\phi_j-1) \oplus S(\phi_j-2)$. For example, in Table 1 we see that the sequence for $\phi=3/5$ is $LSLSL$ while the Fibonacci is $LSLLS$. The reader may wonder why they are different or "reversed". The answer lies in the factor $1/2$ that appears in Eq. (\ref{EqSigma}), this is equivalent to a global phason shift. However, in quasicrystals one needs to compare shifts of a sequence in order to decide if they are or not in the same isomorphism class \cite{NaumisPhason} For example, we can apply several phason shifts to the sequence  $LSLSL$. This is equivalent to an origin shift with cyclic boundary conditions. We
obtain  $LSLSL \rightarrow LLSLS \rightarrow SLLSL \rightarrow LSLLS$. Now the last sequence is the usual Fibonacci sequence and thus both sequences are in the same isomorphism class.

As is well known, an alternative way to generate such sequences is by using either deflation, inflation or recursive rules\cite{Levine,Steurer}. The important result here is that we can relate different fluxes by such deflation/inflation rules. In Fig. \ref{SymbolicButterfly} we explain the previous constructions on the original Hofstadter butterfly.  Fig. \ref{SymbolicButterfly} is meant to be compared with the sequence of Table 1.

Clearly, for other rational sequences as for the silver, bronce, etc. means, one can build such rules, and in fact, the general inflation/deflation rules generated by Eq. (\ref{Eqbands}) have been  extensively  studied 
in the context of quasicrystals \cite{Levine,Steurer}.

In fact, a neat and suggestive way to write the symbolic sequences associated with each $\phi$ is by using $1$ and $0$ instead of $L$ and $S$. This is done by observing that,
\begin{equation}
 S(\phi) =\left[sgn\left (\{\phi  r +\frac{1}{2}\} -\{\phi  (r +1) +\frac{1}{2}\}\right)+1\right]/2
\end{equation}
where $sgn(x)$ is the algebraic sign of $x$ ($+1$ or $-1$) for $x\neq 0$ ($sgn(0)$ is defined as $0$). The previous equation can be interpreted as engineers 
do by looking at $r$ as a continuous variable, say the time, and $S(\phi)$ a square wave with period $\phi$ sampled with frequency one. A dynamical map can be assigned to such sequence,

\begin{eqnarray}\label{circlemap}
  S(\phi)=\left \{ 
    \begin{array}{cc}
      1 & \text{if}\ \{\phi  r +\frac{1}{2}\}<\phi \\
      0 & \text{otherwise} 
    \end{array}\right. 
\end{eqnarray}

\begin{figure*}
 		\includegraphics [width =1.0\linewidth ,height=0.40\linewidth ]{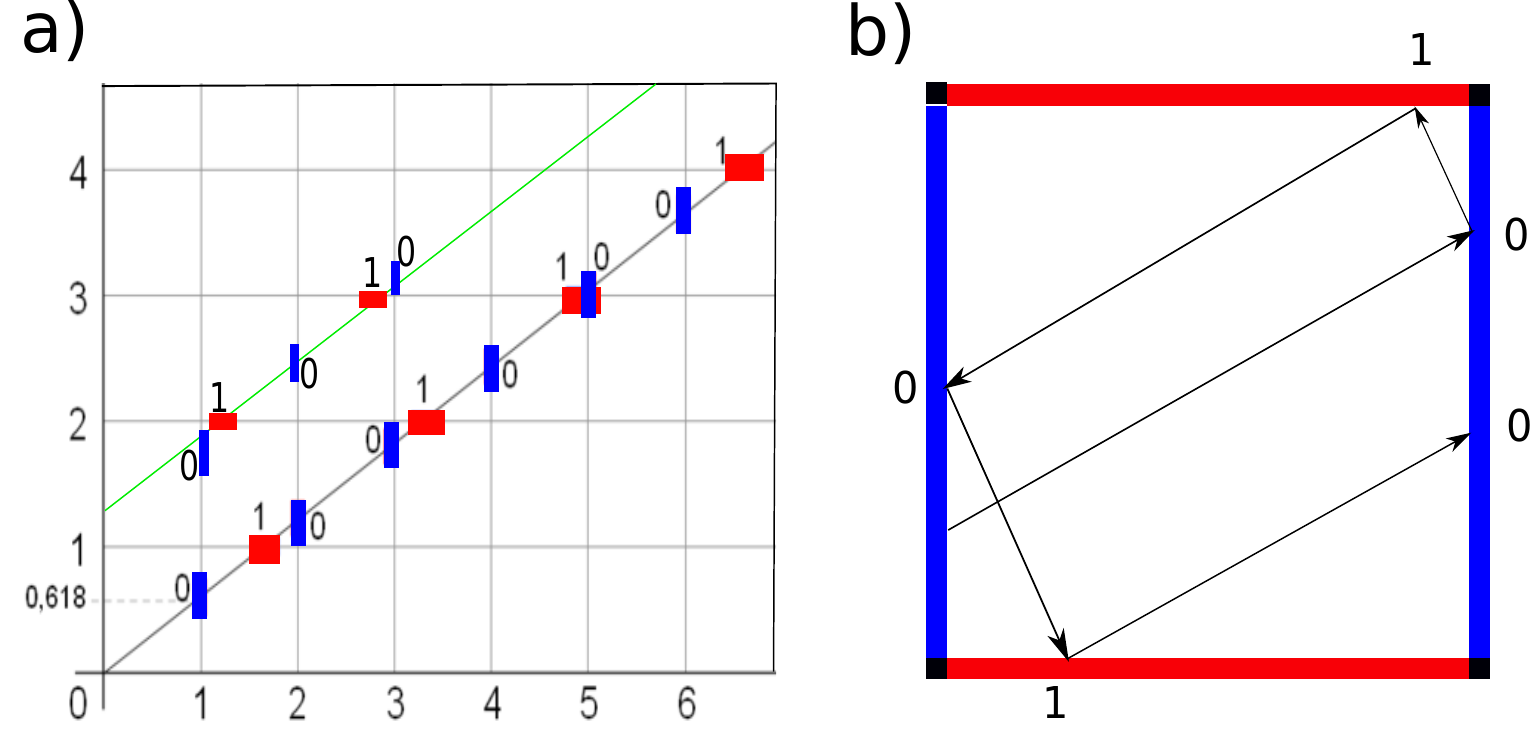}
 		\leavevmode  \caption {Panel a), Sturmian coding of $\phi=3/5$. The black line going through the origin is the usual Sturmian coding for a flux $\phi=3/5=0.618$. This is an approximant of the Golden mean. The slope of the inclined line is $\phi$. Each intersection is labeled as $0$ or $1$ depending on the kind of intersection with the grid. The displaced green line is the same sequence with a global phason shift, and can be compared with the sequence of Table 1. Observe how the color coding is the same as in the Hofstadter butterfly conductances seen at $\phi=3/5$ in Fig. \ref{SymbolicButterfly}. Panel b), the coding can be found in a square colored billiard, in which each kind of reflection, with a vertical or horizontal wall is coded with a $0$ or $1$. The reason is that one can fold the trajectory shown in panel a) by thinking each intersection with the grid as mirrors.} 
 		\label{Sturm}
 	\end{figure*}

Both the symbolic sequence or the dynamical map gives what is called the Sturmian coding of a number \cite{Sturm}, in this case $\phi$. The Sturmian 
coding is an alternative to the continued fraction approach
which is very valuable in order to find good approximants of irrational numbers. The Sturmian coding can be easily visualized by a variant of the cut and projection. Take a square lattice, draw a line with slope $\phi$. As seen in Fig. \ref{Sturm} a), each intersection of this line with the verticals of the square lattice is labeled as a $0$, and each intersection with an horizontal line
is labeled $1$. The labeling of the crossings is the Sturmian coding of $\phi$.
Notice how in Fig. \ref{Sturm} a) the global phason shift discussed before turns out to be very clear. By shifting the line vertically we obtain the green line that produces a global shift of the chain, which is the one that needs to be compared with the Hofstadter butterfly conductance. For irrational $\phi$, the associated sequence is aperiodic and results in a Sturmian word \cite{Sturm}. 
 
 Trajectories on the square lattice can also be warped into a torus or can be seen as a billard in which a particle with constant speed is reflected at the walls \cite{Bedaride}. Each $0$ or $1$ is obtained by recording the collision with horizontal or vertical walls of a trajectory with initial slope $\phi$. Fig. \ref{Sturm} b) indicates such procedure.

 The spectrum of the map defined by Eq. (\ref{circlemap}) is made with discrete frequences $f_{s,l}$ and amplitudes  $\tilde S_l(\phi)$ given by,
 \begin{equation}
 \tilde S_l(\phi)=\frac{sin(\pi l\phi)}{\pi l},  \hspace{6cm}  f_{sl}=s+l\phi
 \end{equation}

\begin{figure*}
 	\includegraphics [width =0.7\linewidth ,height=0.7\linewidth ]{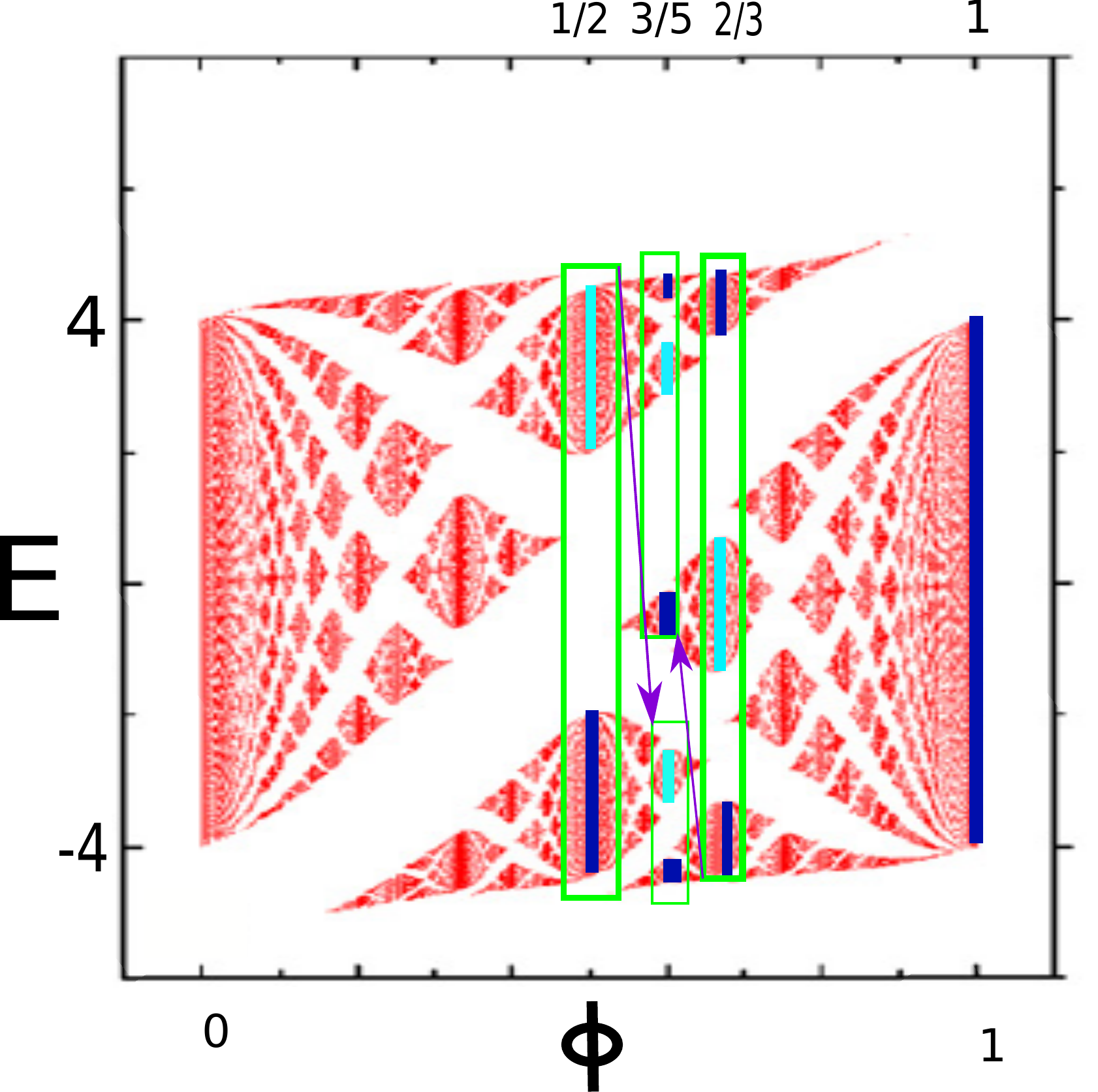}
 	\leavevmode  \caption {The same symbolic sequences for the fluxes seen in Fig. 	\ref{SymbolicButterfly} over imposed on the Fibonacci butterfly, which is obtained by a square well potential \cite{NaumisPHYSB}. Bands with index L are indicated in blue, while bands with S are in light blue. From this figure, the topological equivalence with the Hofstadter butterfly is clearly seen.} 
 	\label{FibonacciButer}
 \end{figure*}

The proposed scheme can also be used to characterize other systems, like the square well potential which contains as a special case, the Fibonacci chain potential \cite{NaumisPHYSB}. As seen in Fig. \ref{FibonacciButer}, this  allows to produce a Fibonacci butterfly . It has been shown that such potential is in the same topological class as the Hofstadter butterfly \cite{QCti}. Thus, all of the methodology developed here can be applied. As an example, in Fig. \ref{FibonacciButer} we indicate the same symbolic sequences for the band conductances seen in 	Fig. \ref{SymbolicButterfly}.

In Figs. \ref{SymbolicButterfly} and \ref{FibonacciButer}, it  is interesting to observe that  band conductances are related to band widths. This can be understood in terms of general arguments concerning the electron's wavefunction overlap in systems that are rational approximants to quasicrystals \cite{NaumisTrace}.  In fact, a dynamical map can be used to investigate the scaling exponents for critical states, and relate them with band-width scaling \cite{NaumisTrace,NaumisJP}.

\section{Methods to calculate Chern numbers and global fractality}\label{Recursion}
 
 From the previous inflation/deflation rules is possible to reverse the procedure, i.e., to obtain the Chern numbers for each gap by a simple recurrence relation. This allows to bypass the need to solve a Diophantine equation. Such procedure is readly obtained from observing that the Chern number for a gap $r$ is the sum of all band conductances up to the given filling fraction \cite{Fradkin},
 \begin{equation}
  \sigma_r=\sum_{s=1}^{r} \sigma_B(s) 
 \end{equation} 
 from where,
 \begin{equation}\label{Eq:ChernRecu}
  \sigma_r=\sigma_{r-1}+\sigma_B(r) 
 \end{equation} 
 It follows that we only need to find the two letter sequence and assign to each letter its numeric counterpart and then sum the sequence at each step. Let us show a simple example. Suppose that we want to calculate the Chern numbers for $\phi=5/8$ without solving the Diophantine equation. We simply use the Fibonacci rule  $S(\phi_j)=S(\phi_j-2) \oplus S(\phi_j-1)$  to produce the sequence $S(5/8)$,
 \begin{equation}
  S(5/8)=LSLLSLSL=-3,5,-3,-3,5,-3.-5,-3
 \end{equation}
where the last step requires the numerical equivalence of a letter, in this case   $S \rightarrow -p $ and $L \rightarrow (q-p)$ as $q-p$ is odd. The sequence of Chern numbers is obtained by using the recurrence relationship Eq. (\ref{Eq:ChernRecu}) and the initial condition $\sigma_0=0$,
\begin{eqnarray}
 \sigma_1=0-3=-3 \rightarrow \sigma_2=-3+5=2 \rightarrow  \sigma_3=2-3=-1 \rightarrow \sigma_4=-1-3=-4 \\
 \sigma_5=-4+5=1 \rightarrow  \sigma_6=1-3=-2 \rightarrow \sigma_7=-2+5=3 \rightarrow \sigma_8=3-3=0.
\end{eqnarray}

A simple comparison with table 1 shows that the sequence is correct and valid for the Hofstadter and Fibonacci butterflies. If the recursion rule for sequence is not known, there are two options. The first is to build the Sturmian coding of $\phi$. The second option is much more efficient: use a simple recursive test.
This option works as follows. Determine the sign $\zeta$. As always $\sigma_{0}=0$, the next Chern number $\sigma_1$ is either $\sigma_0-\zeta p$ or $\sigma_0+\zeta (q-p)$. A direct sustitution in the Diophantine equation gives the right choice. Once $\sigma_1$ is known, $\sigma_2$ can be calculated in a similar way. The method is iterated by using always the previous Chern number as a seed, i.e., $\sigma_{n+1}=\sigma_{n}-\zeta p$  or  $\sigma_{n+1}=\sigma_{n}-\zeta (q-p)$.

 Yet, there is another powerful method to find the Chern numbers. This method reveals several fractal properties of the butterfly. This method is based in the observation made is section \ref{High}
 that solutions are obtained from $\boldsymbol{F}(\phi) \cdot \boldsymbol{T}_1 =1$. With the vector $\boldsymbol{T}_1 =(\sigma_1,\tau_1)$ we define a flux $\phi'=-\tau_1/\sigma_1$, which turns out to be a Farey neighbor of $\phi$. The argument is follows; for two reduced fractions $\phi=p/q$ and $\phi'=p'/q'$, the mediant is defined as \cite{Schroeder},
 \begin{equation}
  \frac{p''}{q''} = \frac{p+p'}{q+q'}
 \end{equation}
This requires the fractions to be unimodular $|p'q-pq'|=1$. Such construction is easily understood in  two dimensions, as,
\begin{equation}\label{Eq:Vecsum}
 \boldsymbol{F}(\phi'')=\boldsymbol{F}(\phi)+\boldsymbol{F}(\phi')
\end{equation}
 and the condition for unimodularity is $\boldsymbol{F}(\phi) \cdot \boldsymbol{F}^{\perp}(\phi')=\pm 1$. Thus we can identify the fundamental solution as,
\begin{equation}
 \boldsymbol{T}_1 = \boldsymbol{F}^{\perp}(\phi')  \hspace{1cm} \text{with} \hspace{1cm} \phi'=-\tau_1/\sigma_1
\end{equation}

 Mediants occurs naturalley in Farey sequences, defined as fractions between $0$ and $1$ of a given largest denominator \cite{Schroeder}. In this sequence, each fraction is the median of its two neighbors.  Thus, we just proved that given a flux $\phi$, the fundamental solution is given by one of the Farey neighbors. Let us workout an example to reproduce some results of Table 1. Consider the Farey sequence of order 5 built from a Farey tree \cite{Schroeder},
 \begin{equation}
  \frac{0}{1}, \ \frac{1}{5}, \ \frac{1}{4}, \ \frac{1}{3}, \ \frac{2}{5}, \ \frac{1}{2}, \ \frac{3}{5}, \ \frac{2}{3}, \ \frac{3}{4}, \ \frac{4}{5}, \ \frac{1}{1}
 \end{equation}
 and apply it to the flux $\phi=2/3$. Its upper Farey neighbor $\phi'=3/4$ satisfies $\boldsymbol{F}(2/3) \cdot \boldsymbol{F}^{\perp}(3/4)=(2,3)\cdot (-4,3)=1$ as expected for a Farey sequence. Thus we identify $\boldsymbol{F}^{\perp}(3/4)=\boldsymbol{T}_1=(-4,3)$. The solution can be folded back to the "Chern first Brillouin zone" by taking the modulus with $\boldsymbol{F}^{\perp}(2/3)$ as explained in section \ref{High}, from where  $\boldsymbol{T}_1=(-4,3)+(3,-2)=(-1,1)$ resulting in $\sigma_1=-1$, coinciding with Table I.  As a matter of fact, each flux in the Farey sequence provides all its own solutions by using its right neighbor fraction in the sequence. Its is important to remark that the approximants of the golden ratio, given by the Fibonacci numbers, are also Farey neighbors \cite{Schroeder}. 
 
 These observations suggest the possibility to understand the self-similarity of the topological phase diagram by observing how phases are related at different fluxes. Indeed this is the case. Consider the vectorial sum Eq. (\ref{Eq:Vecsum}) applied to the product,
 \begin{equation}\label{Eq:FT}
  \boldsymbol{F}(\phi'') \cdot \boldsymbol{T}_1(\phi)=[\boldsymbol{F}(\phi)+\boldsymbol{F}(\phi')]\cdot \boldsymbol{T}_1(\phi)
 \end{equation}
 where now we changed the notation to indicate that $\boldsymbol{T}_1(\phi)$ is a fundamental solution for flux $\phi$, i.e.,$\boldsymbol{F}(\phi) \cdot \boldsymbol{T}_1(\phi)=1$, {\it before doing the folding using the vector  $\boldsymbol{F}^{\perp}(\phi)$}. Using this fact and that $\boldsymbol{F}(\phi')\cdot \boldsymbol{T}_1(\phi)=\boldsymbol{F}(\phi')\cdot \boldsymbol{F}_1^{\perp}(\phi')$, it follows that,
 \begin{equation}\label{Eq:Solphiprime}
   \boldsymbol{F}(\phi'') \cdot \boldsymbol{T}_1(\phi)=1
 \end{equation}
 proving that  $\boldsymbol{T}_1(\phi)$ is a fundamental solution of the mediant $\phi''$ obtained from $\phi$ and $\phi'$. It is important to remark that the solution can be folded to have Cherns between $q/2$ and $-q/2$ by using the rule $ \boldsymbol{T}_1(\phi)-s\boldsymbol{F}^{\perp}(\phi'')$ for some integer $s$.
 
 Consider as an example the same flux $\phi=2/3$, with Farey neighbour $\phi'=3/4$. The resulting mediant is $\phi''=(2+3)/(4+3)=5/7$. We have that,
 \begin{equation}
  \boldsymbol{F}(5/7) \cdot \boldsymbol{T}_1(2/3)=(5,7)\cdot (-4,3)=1
 \end{equation}
 as predicted. By folding back by $\boldsymbol{F}^{\perp}(5/7)=(-7,5)$ it gives the fundamental solution $(3,-1)$, i.e., the first Chern for $\phi=5/7$ is $3$. 
 It is important to remark that if we use the unfolded solution $\boldsymbol{T}_1(\phi)=(-1,1)$, instead of $(-4,3)$, we will not get the fundamental solution but a shifted one. This comes out as follows, let us consider again the product (\ref{Eq:FT}) but with a folding,
    \begin{equation}
   \boldsymbol{F}(\phi'') \cdot (\boldsymbol{T}_1(\phi)+s\boldsymbol{F}^{\perp}(\phi))=1+
   s\boldsymbol{F}(\phi') \cdot \boldsymbol{F}^{\perp}(\phi)
   \end{equation}
   The product $\boldsymbol{F}(\phi') \cdot \boldsymbol{F}^{\perp}(\phi)$ although being integer, is not zero in general, resulting in a solution different from $r=1$.

 What is remarkable about Eq. (\ref{Eq:Solphiprime}) is that the sequence for a flux $\phi''$ is contained and generated by the same solution than $\phi$, as we can simply multiply  Eq. (\ref{Eq:Solphiprime}) by $r$,
 \begin{equation}
   \boldsymbol{F}(\phi'') \cdot [r\boldsymbol{T}_1(\phi)]= \boldsymbol{F}(\phi'') \cdot [\boldsymbol{T}_r(\phi)]=r
 \end{equation}

 where $\boldsymbol{T}_r(\phi)$ is the solution for gap $r$ for a $\phi$ which is above $\phi''$ in the Farey tree. Yet the folding is dictated by $\phi''$ instead of $\phi$. As a matter of fact, it means that we were able to find a construction based in blocks of sequences as happens with the Fibonacci ones, but this time, for any rational flux, as the Farey tree will eventually contain the fraction. Such construction can be seen in Figs. \ref{SymbolicButterfly} and \ref{FibonacciButer} for the fractions $1/2, 3/5, 2/3$. This helps to explain the previously numerically observed relationships between the global fractality of the butterfly and Farey neighbors sequences, as well as for its representation as Ford circles \cite{InduEPJ,Indubook}.

\section{Harper potential and the cut and projection method}\label{diffraction}

One may wonder what is behind the fact that the cut and projection method predicts the topological phases. The reason is that band gaps where topological modes reside, are open due to electron diffraction, as stationary waves are produced when the wavevector $\boldsymbol{k}$ is equal to a reciprocal lattice vector $\boldsymbol{Q}$. Thus, a vanishing group velocity $\boldsymbol{v}_{g}(\boldsymbol{k})$ is observed and a Van Hove singularity occurs.  Formally, band gaps and diffraction are related through the general formula for the density of states $\rho(\epsilon)$,
\begin{equation}
\rho  (\epsilon) = \oint _{\boldsymbol{S}(\epsilon)}\frac{d\boldsymbol{S}}{2 \pi^{2} \vert \boldsymbol{v}_{g}(\boldsymbol{k}) \vert }
\end{equation}
where $\epsilon$ is the energy and $\boldsymbol{k}$ the wavevector. The integral is made along contours $\boldsymbol{S}(\epsilon)$ of equal energy. 
The group velocity is determined by the energy dispersion $\boldsymbol{v}_{g}(\boldsymbol{k})=\nabla _{\boldsymbol{k}}\epsilon (\boldsymbol{k})$. Whenever diffraction occurs, $\vert \boldsymbol{v}_{g}(\boldsymbol{k}) \vert=0$. The previous formula explains the Van Hove logarithm singularities and related topological collisions at each Hofstadter butterfly  band center \cite{Naumis}.

Let us now understand how the cut and projection method is related with bands.  We start our analysis by using the identity $x=\lfloor x\rfloor+\{x\}$ applied to $m \phi+\nu_y$ in the Harper potential given in Eq. (\ref{EqHarper}), 
\begin{equation}
 V(m)=2\lambda \cos( 2\pi \{ m \phi +\nu_{y}\})
\end{equation}
 
Next we observe that lower band edges are obtained by seeting $\nu_{y}=0$ 
in Eq. (\ref{EqHarper}). The other limiting value $\nu_{y}=1/2$  gives the upper band edges \cite{Thouless}. As we are only interested in states at band edges, in whatfollows we will only consider lower band edges $\nu_y=0$ since upper band edges
share the same Chern numbers as the contiguous lower band edge. In such case, 
using Eq. (\ref{map}), we can reinterpret $m$ as a Chern number, i.e., $m=\sigma_r$,
from where the fractional part can be associated with the band index,    
\begin{equation}\label{EqChernOrdering}
 V(\sigma_r)=2\lambda \cos( 2\pi \{ \sigma_r \phi \}q/q)=2\lambda \cos(2\pi r /q)
\end{equation}

Now is clear how the argument of the cosine is associated with a wave-vector  $k=2\pi r/q$, having $r=0,...,q-1$. Furthermore, 
using Eq. (\ref{EqChernOrdering}) and Eq. (\ref{EqXperpFilling}), it follows that,
\begin{equation}\label{EqChernOrdering2}
 V(\sigma_r)=2\lambda \cos( 2\pi \sqrt{p^{2}+q^{2}} |\boldsymbol{X}_{\sigma_r,\tau_r}^{\perp}|)
\end{equation}
Also, as $|\boldsymbol{X}_{\sigma_r,\tau_r}^{\perp}|< 1$, this shows that $V(\boldsymbol{\sigma_r})$ induces an ordering of the potential according to its distances
in $E^{\perp}$. Since band-level crossings do not  happen \cite{Fradkin}, the ordering is preserved for all $\lambda$. Alternatively, we can say that
ordering is provided by the Chern number map of Eq. (\ref{EqXperpFilling}).


 Let us explain in detail the previous assertion. Following Fradkin \cite{Fradkin},
consider the limit $\lambda \rightarrow \infty$. Then $\psi_m^{r} \approx \delta(m-m_r)$, i.e., the wave function is a delta centered at some site $m_r$ for band $r$. To find where is it localized, from Eq. (\ref{EqHarper}) this
will happen whenever the energy of the level is,
\begin{equation}
 E_r  \approx V(m_r)=2\lambda \cos( 2\pi \{m_r \phi \}q/q)
\end{equation}
Setting $m_r=\sigma_r$ we obtain, 
\begin{equation}
E_r \approx 2\lambda \cos (2 \pi r/q)=2\lambda \cos( 2\pi \sqrt{p^{2}+q^{2}}|\boldsymbol{X}_{\sigma_r,\tau_r}^{\perp}|)
\end{equation} 

The process can be 
summarized as follows. For a band $r$, the state is localized at site $m_r=\sigma_r$. Or in an alternative way, given a site $m_r=\sigma_r$, its associated band position is determined  by $|\boldsymbol{X}_{\sigma_r,\tau_r}^{\perp}|$. 

Notice that due to the parity of $V(m)$, the localization can also
happen at $m=-\sigma_r$ for the same energy. Since for a rational $\phi$ the lattice is periodic, $m$ needs to be folded back into sites $m=0,...,q-1$. 
By performing the right folding depending wheter $q$ is odd or even, one finds  that the delta functions are separated by a Chern number of sites, and results in the
Chern phenomena beating discovered in Ref. \cite{InduPRL}. This phenomena implies that edge states for each band are a convolution of the Chern doublets
with the ground state. For the case of $\lambda=1$, the doublet is convoluted with a fractal ground state resulting in fractal doublets\cite{InduPRL}. 

Also, Aubry and Andre proved that the Harper equation 
is self-reciprocal \cite{Harper}, i.e, the Fourier coefficients of the wave-function follow the same Harper equation but with $\lambda$ 
replaced using the rule $\lambda \rightarrow 1/\lambda$. As a result, for $\lambda \rightarrow 0$, the Fourier coefficients of the wave-function are just 
delta-localized at $k=\pm \sigma_r$ resulting in the wave-function,  
\begin{equation}
 \psi_m^{r} \approx \frac{1}{q} cos(2 \pi \sigma_r m)
\end{equation}
for the band $r$. 

Here we proved that the map $r=\{ \sigma_r \phi \}q$ allows to order the energies in terms of the potential $V(\sigma_r)$, and this ordering
is the same as the one for the wave-function Fourier coefficients. We can also reinterpret Eq. (\ref{EqChernOrdering})  in the original 
framework proposed by Hofstadter, i.e. the Bloch-Floquet theorem for the  wave-function in real space leads to just a re-ordering 
in reciprocal space for the original wave-functions \cite{Hof}. The order is dictated by the perpendicular component of $\boldsymbol{X}$.

It is worthwhile mentioning that around a given flux,  several topological sequences can be obtained by tilting \cite{Naumis} $\phi$ by a small amount $\delta \phi$. This is equivalent to introduce phason disorder, and as a consequence, the resulting sequences have satellites in the diffraction pattern \cite{NaumisPhason}. Similar patterns are observed on graphene over a sustrate \cite{Taboada,TaboadaRipple}.

\section{Conclusions}
Using ideas from quasicrystals,  in particular the cut and projection method, we were able to find several interesting properties of the Diophantine equation which characterizes
the Hofstadter butterfly as a topological phase diagram. We showed that the bands conductance for any given rational flux are described by symbolic sequences. Thus, bands 
conductance  at different fluxes can be related by inflation/deflation rules as happens for rational approximants of quasiperiodic sequences. Such rules correspond to the Sturmian sequence of the flux. They can be obtained by using a dynamical map, a trajectory in a torus or in a square billiard, resulting in easy rules to find Chern numbers.  The presented mechanism is also valid for the square well potential which leads to the Fibonacci butterfly  \cite{NaumisPHYSB}. We also have a higher dimensional construction that allows to find solutions and its self-similarity through Farey sequences, trees and neighbours.
This allows to describe topological phases within the context of quasicrystals,
which is seems to be useful in order to describe complex phases in Moire patterns of graphene over graphene at magical angles \cite{Tarnopolsky2019}.  \\

This work has been supported by UNAM-DGAPA project IN102717.

\end{document}